\documentclass[onecolumn,showpacs,prd]{revtex4}
\usepackage{epsfig}
\usepackage{bm}
\usepackage{amsfonts}
\usepackage{amssymb,amsmath}
\usepackage{enumerate}
\usepackage{multirow}
\usepackage{epsfig}
\usepackage{graphicx}
\usepackage{bm}
\usepackage{hyperref}

\begin{document}

\title{A field theoretic approach to the energy momentum tensor for theories coupled with gravity}
\author{Pradip Mukherjee}
\email{mukhpradip@gmail.com}
\affiliation{Department of Physics, Barasat Govt. College, 10 K. N. C. Road, Barasat, Kolkata 700 124, India}
\altaffiliation{Visiting Associate, Inter University Centre for Astronomy and Astrophysics, Pune, India}
\author{Anirban Saha}
\email{anirban@associates.iucaa.in}
\affiliation{Department of Physics, West Bengal State University, Barasat, North 24 Paraganas, West Bengal, India}
\altaffiliation{Visiting Associate, Inter University Centre for Astronomy and Astrophysics, Pune, India}
\author{Amit Singha Roy}
\email{singharoyamit@gmail.com}
\affiliation{Department of Physics, Barasat Govt. College, 10 K. N. C. Road, Barasat, Kolkata 700 124, India}
\begin{abstract}
We provide a field-theoretic algorithm of obtaining energy momentum tensor (EMT) for gravitationally coupled scalar field theories. The method is equally applicable to both minimal and non-minimal coupling. The algorithm illuminates the connection between the EMT, obtained by functional variation of the metric, and local balance of energy and momentum.
\end{abstract}
\maketitle
In the original formulation of general relativity (GR), Einstein introduced the physical energy momentum tensor (EMT) of the matter theory as the source of curvature \cite{E} and curvature in turn determined the motion of the source. Subsequently, Hilbert enunciated an action principle to obtain Einstein's equations in the usual way, i.e. by extremizing an action. Various prescriptions for obtaining an EMT were developed \cite{Landau, AE, Weinberg}. The most popular algorithm \cite{Weinberg} is to vary the action with respect to the background metric, leading to the definition
\begin{equation}
\Theta^{\mu\nu} = - \frac{2}{\sqrt{-g}}\frac{\delta S_\Phi}{\delta g_{\mu\nu}}
\label{def}
\end{equation}
where, $S_\Phi$ is the generic action of the source fields minimally coupled to external gravity
 The EMT thus obtained is symmetric and covariently conserved. Physics of interactions other than gravity are described by field theories in the flat (Minkowskian) tangent space.Equation (\ref{def}) can still be used to construct a symmetric EMT by coupling the theory minimally with external gravity and finally making the metric flat.However definition (\ref{def}) has one disadvantage -- its connection with energy and momentum flow is not much apparent.
 
 There is a well known field theoretic method of constructing an EMT due to Noether who proved \cite{Noether} that corresponding to a continuous symmetry of the field theoretic action there exists a conserved current. Specifically, translational symmetry leads to a conserved EMT \cite{MTW}. 
 In the EMT obtained by Noether's method the connection with energy and momentum flow is strong because here the conservation of EMT is connected with the translational symmetry of the system and the energy momentum four-vector is generator of translation. 
 Note that in Nother's approach both construction of EMT and its conservation is valid on shell whereas the construction of the symmetric EMT (\ref{def}) is apparently off shell. However, the conservation of the latter depends on the equation of motion. So essentially
 the EMTs are defined in both the approaches modulo equation of motion. There are examples where the two methods give the same result. Such comparisions has so far been done for minimally coupled theories.For non-minimally coupled theories the definition (\ref{def}) can not be applied as such. The usual way out is to write the equation of motion corresponding to the metric $g_{\mu \nu}$ and read-off the EMT by rearranging different terms therein. Naturally there is ambiguity as the rearrangement follows no definitive prescription \cite{PB, torre, H, S}. But it can be shown that the differenet definitions are equivalent using the equations of motion. To our knowledge, a Noether algorithm to construct the ETM of a scalar matter coupled with gravity is not available in the literature. We provide such an approach, applicable to both minimally and non-minimally coupled scalar field theories, in the present paper. 
 
The essence of our approach consists of the following. We follow the spirit inherent in (\ref{def}) which is the response of the source due to variation of the background metric where the metric is external and not dynamical. As is well-known the primitive definition of EMT is based on the balance of energy and momentum which is a local phenomena. Hence, to construct an EMT from the physical perspective the locally inertial frame is particularly suitable. A choice of Lorentzian coordinates may be assumed which will henceforth be denoted by adapted coordinates. In these coordinates, the metric assumes value as in the flat (Minkowski) space time and its first derivtives vanish. Nevertheless the curvature of the background spacetime shows up in the nonvanishing second derivatives of the metric. The crux of our method is to view the corresponding action as an auxiliary field theory in the a flat Minkowski spacetime which has an  infinitesimal overlap with the tangent space at P.
The whole process can be summarised in the following algorithm:
\begin{enumerate}
\item 
\label{step1}
From the original theory subtract the dynamic part for pure gravity ( i.e. Einstein--Hilbert action). This will give the scalar field theory interacting with external gravity.

\item
\label{step2}
Express the resulting action in the adapted coordinates, at a point P, which have the properties
\begin{eqnarray} 
g_{\mu\nu} & = & \eta_{\mu\nu} = {\rm diag}\left(-1, 1, 1, 1 \right)
\label{g} \\
\partial_\lambda g_{\mu\nu} & = & 0
\label{delg} \\
\partial_\lambda \partial_\rho g_{\mu\nu} & \ne & 0
\label{del2g}
\end{eqnarray}
Note carefully that the conditions (\ref{g}, \ref{delg} and \ref{del2g}) hold in the curved space-time within the infinitesimal patch around point P, in locally inertial frame. Uptill now $g_{\mu \nu}$ refers to the metric of the curved space-time.
\item
\label{step3}
The action obtained in step (\ref{step2}) will now be considered as a new field theory in the Minkowskian space which has an  infinitesimal overlap with the tangent space at the point P, with metric $\eta_{\alpha \beta}$. $\phi$ and $g_{\mu \nu}$ are now respectively scalar and second rank tensor fields in this space. This will be henceforth referred to as the auxiliary field theory. To avoid confusion we will denote the second rank tensor field by $H_{\mu\nu}$ instead of $g_{\mu\nu}$. Obviously, only second derivatives of $H_{\mu \nu}$  will appear in this theory. Note that the dynamics of this newly defined second rank tensor field $H_{\mu\nu}$ is {\it{not}} subject to the conditions (\ref{g}, \ref{delg}, \ref{del2g}). One should always remember that within the perview of the auxiliary field theory $H_{\mu\nu}$ has nothing to do with the metric either of the curved space-time or of the Minkowskian space-time.

\item
\label{step4}
Now compute the EMT by applying Noether's theorem using the translation symmetry of the new theory. Since the energy-momentum four vector is a generator of translation symmetry the EMT actually corresponds to the balance of energy and momentum all over the flat space including the tangent space at point P.

\item
\label{step5} The auxiliary field theory becomes identical with that obtained in (\ref{step2}) in the overlap if the fields $H_{\mu\nu}$ are replaced by $g_{\mu \nu}$ and the constraints (\ref{g}, \ref{delg}, \ref{del2g}) are imposed. From steps (\ref{step1}), (\ref{step2}) and (\ref{step3}) it can be understood that the EMT obtained in step (\ref{step4}) is same as the EMT of the original scalar field theory (\ref{QA}) interacting with gravity locally at a point P in the region of overlap. 

\item
\label{step6}
The final task is to express the EMT in step(\ref{step5}) in terms of general coordinates in curved space-time. Care should be taken in this step so that all the terms that appear in this EMT has unambiguous geometric meaning. What we mean by the phrase 'unambiguous geometric meaning' is clarified in the following,see below equation(\ref{T11}). 

\item
\label{step7}
The EMT thus obtained should serve as the source for the gravitational field in the original theory.
\end{enumerate}

From the above description of the proposed method it is apparent that the procedure is applicable for a generic scalar field theory coupled to gravity. For definiteness we take a non-minimally coupled quintessence model to illustrate our method though the same algorithm is applicable in principle for different dynamics of the scalar field, such as k-essence.
\begin{figure}[h]
\begin{center}
\end{center}
\includegraphics{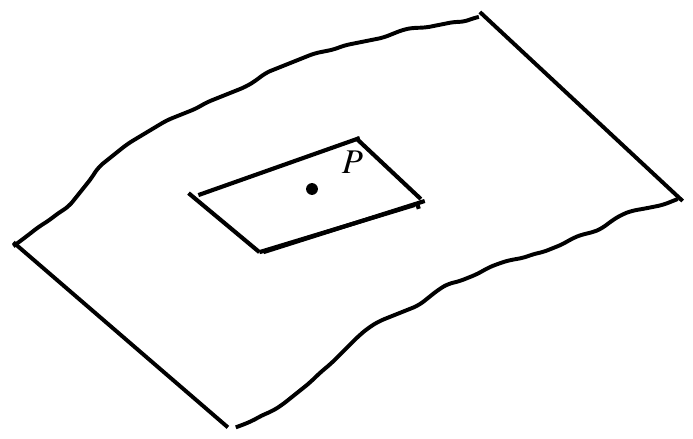}
\caption{ Field theory defined on the tangent space}
\end{figure}

We start with the following action:
\begin{equation}
S=\int{{d^4}x}{\sqrt{-g}}\left[\frac{{M_{pl}^2}}{2}(1-F(\phi))R-\frac{1}{2}{g^{\mu\nu}}({\nabla_{\mu}}\phi)({\nabla_{\nu}}\phi)-V(\phi)\right]
\label{QA}
\end{equation}
Note that a non-zero $F(\phi)$ signifies nonminimal coupling. In (\ref{QA}), ${M_{pl}^2}$ is given by $\left( 8\pi G\right)^{-1}$.

As the first step of our procedure we abstract from (\ref{QA}) the form of the theory coupled with curved space-time when the metric is external. This leads to the Lagrangian
\begin{eqnarray}
{\mathcal{L}_{\phi}}=- \sqrt{-g} \left[ \frac{{M_{pl}^2}}{2}F(\phi)R + \frac{1}{2}{g^{\mu\nu}}({\nabla_{\mu}}\phi)({\nabla_{\nu}}\phi) + V(\phi) \right]
\label{L}
\end{eqnarray}
Note that metric in (\ref{L}) is a background field and influences the motion through the coupling with the scalar field through curvature.

So far, the coordinate system was general, charting the curved spacetime. The next step of our method is to concentrate at the neighbourhood of the point P (see fig.1) and use the adapted coordinates. 
 The Riemann tensor is expressed, using (\ref{g}), (\ref{delg}), in the form \cite{Carroll},
\begin{eqnarray}
R_{\alpha\beta\gamma\delta} = \frac{1}{2}\left({g_{{\beta\gamma},{\alpha\delta}}}-{g_{{\alpha\gamma},{\beta\delta}}}-{g_{{\beta\delta},{\alpha\gamma}}}+{g_{{\alpha\delta},{\beta\gamma}}}
\right)\label{1}
\end{eqnarray}
\noindent The Ricci tensor is 
\begin{eqnarray}
R_{\beta\delta} = \eta^{\alpha\gamma}R_{\alpha\beta\gamma\delta} 
\label{Ricci} 
\end{eqnarray}
The Ricci scalar is obtained on another contraction as,
\begin{eqnarray}
R &=& {\eta^{\alpha\gamma}}{\eta^{\beta\delta}}{R_{\alpha\beta\gamma\delta}} = -\left(\eta^{\alpha\gamma}\partial_\lambda \partial^\lambda{ g_{\alpha\beta}} -\partial^\alpha \partial^\beta g_{\alpha\beta}
 \right)\label{2}
\end{eqnarray}

The Lagrangian (\ref{L}) can now be written in terms of the adapted coordinates, using (\ref{2}) as 
\begin{eqnarray}
\mathcal{L} = \frac{{M_{pl}^2}}{2} F\left(\phi \right)\left\{\eta^{\alpha\gamma}\partial_\lambda \partial^\lambda{ g_{\alpha\beta}} -\partial^\alpha \partial^\beta g_{\alpha\beta} \right\} -\frac{1}{2}{\eta^{\mu\nu}}({\partial_{\mu}}\phi)({\partial_{\nu}}\phi) - V(\phi)
\label{flat_L}
\end{eqnarray}
Note that due to the condition (\ref{g}), $\sqrt{-g} = 1$.
The region in which the adapted coordinates are defined is the tangent space at the point $P$.
The range of these can however, be extended to the whole flat (Minkowski)space that has a  region of overlap with the tangent space at P.

As the next step of our algorithm we will construct the Lagrangian of the auxiliary field theory in these coordinates as
\begin{eqnarray}
\mathcal{L} = \frac{{M_{pl}^2}}{2} F\left(\phi \right)\left\{\eta^{\alpha\gamma}\partial_\lambda \partial^\lambda{ H_{\alpha\beta}} -\partial^\alpha \partial^\beta H_{\alpha\beta} \right\} -\frac{1}{2}{\eta^{\mu\nu}}({\partial_{\mu}}\phi)({\partial_{\nu}}\phi) - V(\phi)
\label{auxflat_L}
\end{eqnarray}
where $H_{\mu\nu}$ is a second rank tensor field (see the discussion in (\ref{step3})). These fields, along with the scalar field $\phi$  form a closed system. The theory has translation symmetry (as a part of the more general Poincare symmetry). Note that no constraint is imposed on 
$H_{\mu\nu}$.  
 
The equation of motion for the $\phi$ field is obtained as usual,
\begin{eqnarray}
\Box \phi - \frac{{M_{pl}}^2}{2}F^\prime R_H - V^\prime = 0
\label{eqmphi}
\end{eqnarray}
In the above, $A^\prime$ denotes differentiation of $A(\phi)$ with respect to $\phi$. $ R_H$ is given by 
\begin{eqnarray}
R_H =  -\left(\eta^{\alpha\gamma}\partial_\lambda \partial^\lambda{ H_{\alpha\beta}} -\partial^\alpha \partial^\beta H_{\alpha\beta}
 \right)\label{2H}
\end{eqnarray}
A characteristic feature in the Lagrangian (\ref{auxflat_L}) is the presence of second derivatives of the tensor field $H_{\mu\nu}$. Thus it is a higher derivative field theory \cite{ostro}. The equation of motion for the $H_{\mu \nu}$ field is \cite{git},
\begin{eqnarray}
{{\partial}_\mu}{{\partial}_\nu}\left[\frac{{\partial}\mathcal{L}}{{\partial}({{\partial}_\mu}{{\partial}_\nu}{H_{\alpha\beta}})}\right]-{{\partial}_\mu}\left[\frac{{\partial}\mathcal{L}}{{\partial}({{\partial}_\mu}{{H_{\alpha\beta}})}}\right]+\frac{{\partial}\mathcal{L}}{{\partial}{H_{\alpha\beta}}}=0
\label{eqmg}
\end{eqnarray}
Using (\ref{auxflat_L}), we get the explicit form,
\begin{eqnarray}
{{\partial}_\mu}{{\partial}_\nu}\left[\frac{{M_{pl}}^2}{2}F(\phi)\right]\{\eta^{\alpha\nu}{\eta^{\mu\beta}}-\eta^{\mu\nu}{\eta^{\alpha\beta}}\}=0
\label{eqng}
\end{eqnarray}
Apprently this equation makes no reference to the field $H_{\mu \nu}$. 
However, eliminating $F(\phi)$ from equation(\ref{eqng}) and (\ref{eqmphi}), the explicit equation of motion for the fields $H_{\mu \nu}$ can be obtained. Also one can alternatively write equation (\ref{eqng}) as,
\begin{eqnarray}
\partial^\alpha \partial^\beta F - \eta^{\alpha\beta}\Box{F} =0
\label{f}
\end{eqnarray}

The construction of the EMT, $T^{\mu\nu}$ will now be detailed. Following Noether's prescription,
\begin{eqnarray}
T^{\mu\nu}&=&\frac{{\partial}\mathcal{L_H}}{{\partial}({{\partial}_{\mu}}\phi)}({{\partial}^\nu}\phi)-{{\eta}^{\mu\nu}}
\mathcal{L_H}+\frac{{\partial}\mathcal
{L_H}}{{{\partial}({\partial}_\lambda}{{\partial}_\mu}{H_{\alpha\beta}})}({\partial_\lambda}{\partial^\nu}{H_{\alpha\beta}})-
\partial_\lambda\left( \frac{{\partial}\mathcal{L_H}}{{{\partial}}{{\partial}_\lambda}{{\partial}_\mu {H_{\alpha\beta}}}}\right)({\partial^\nu}{H_{\alpha\beta}})
\label{EMT}
\end{eqnarray}
which gives 
\begin{eqnarray}
T^{\mu\nu} & = &-({\partial^\mu}\phi)({\partial^\nu}\phi)+{{\eta}^{\mu\nu}}\left[\frac{{M_{\rm pl}^2}}{2}F(\phi)R_H+\frac{1}{2}{{\eta}^{\alpha\beta}}({\partial_{\alpha}}\phi)({\partial_{\beta}}\phi) + V(\phi)\right] \nonumber\\
&& - \left[\frac{{M_{\rm pl}}^2}{2}\left(\eta^{\lambda\mu}\eta^{\sigma\beta}-\eta^{\sigma\mu}\eta^{\lambda\beta}
\right) \partial^{\nu}\partial_{\sigma} H_{\lambda\beta}\right]F(\phi)\nonumber\\
&& + \quad\partial_\sigma\left[\frac{{M_{\rm pl}}^2}{2}\left(\eta^{\lambda\mu}\eta^{\sigma\beta}-\eta^{\sigma\mu}\eta^{\lambda\beta} \right)F(\phi)\right]{\partial^\nu}{H_{\lambda\beta}}
\label{T}
\end{eqnarray}
An explicit check of the conservation of $T^{\mu\nu}$ is due. A straight forward calculation shows,
\begin{eqnarray}
{\partial_{\mu}}{T^{\mu\nu}}=0
\label{conserve}
\end{eqnarray}
In arriving at the above conservation law we have used the equations of motion (\ref{eqmphi}) and (\ref{eqng}). Now that we have obtained a conserved EMT for the field theory (\ref{auxflat_L}). For reasons that will be clear in the following, we rewrite (\ref{T}) such that first derivative of $H_{\mu\nu}$ is not explicit in the EMT. It follows as,
\begin{eqnarray}
T^{\mu\nu} & = & - \left(\partial^{\mu}\phi\right) \left(\partial^{\nu}\phi \right) + \eta^{\mu\nu} \left[\frac{{M_{\rm pl}^2}}{2}F(\phi) R_H + \frac{1}{2}\eta^{\alpha\beta}(\partial_{\alpha}\phi)(\partial_{\beta} \phi) + V(\phi)\right]\nonumber\\
&& - \frac{{M_{\rm pl}}^2}{2}\left(\eta^{\lambda\mu}\eta^{\sigma\beta} -\eta^{\sigma \mu}\eta^{\lambda\beta} \right) \left(\partial^{\nu}\partial_{\sigma} H_{\lambda\beta}\right) F(\phi)\nonumber\\
&& + \frac{{M_{\rm pl}}^2}{2}\left(\eta^{\lambda\mu}\eta^{\sigma\beta} - \eta^{\sigma\mu}\eta^{\lambda\beta}\right)\partial^{\nu}\left\{ H_{\lambda\beta}\partial_{\sigma} F(\phi)\right\}
- \frac{{M_{\rm pl}}^2}{2}\left[\eta^{\lambda \mu} \left(\partial^{\beta} \partial^{\nu}F \right) H_{\lambda \beta}  - \eta^{\lambda \beta}\left(\partial^{\mu} \partial^{\nu}F \right) H_{\lambda \beta}\right]
\label{T11}
\end{eqnarray}
The program now is to import the EMT (\ref{T11}) to the local patch at P with substitution of $H_{\mu\nu}$ by $g_{\mu\nu}$ and express it in terms of general coordinates. But for the latter, we have to ensure that all the terms have unambiguous correspondence with geometric objects. For example, the first and second term of (\ref{T11})
(on substitution of $H$ by $g$, are already in a form that has such a correspondence, but not the third term, owing to the presence of the factor $\left(\partial^{\nu}\partial_{\sigma} g_{\lambda\beta}\right)$. We have to improve the EMT (\ref{T11}) so as to get rid of such ambiguity.

A possible way is to add a term $\partial_{\sigma}M^{\sigma\mu\nu}$ to the EMT (\ref{T11}), where $M^{\sigma\mu\nu}$ is a third rank tensor antisymmetric in its first two indices such that $\partial_{\mu}\partial_{\sigma}M^{\sigma\mu\nu}$ vanishes identically. This ensures the conservation of the improved EMT \cite{BF}. An appropriate choice is 
\begin{eqnarray}
M^{\sigma\mu\nu} = \eta^{\nu\beta}\left(\eta^{\mu\lambda} \eta^{\rho\sigma } - \eta^{\mu\rho}\eta^{\lambda \sigma}\right)\left[ \left(\partial_{\rho} H_{\lambda\beta} \right) F(\phi) - 3 \left\{\partial_{\rho}F(\phi) \right\}  H_{\lambda\beta}  \right]
\label{M}
\end{eqnarray}
Note that $M^{\sigma\mu\nu}$ has the required anti symmetry  in $\sigma $ and ${\mu}$.
Adding $\frac{{M_{\rm pl}}^2}{2}\partial_\sigma M^{\sigma\mu\nu}$ with $T^{\mu\nu}$ to (\ref{T11}), we get the improved tensor,
\begin{eqnarray}
\Theta^{\mu\nu} & = &-(\partial^{\mu}\phi)(\partial^{\nu}\phi) + \eta^{\mu\nu}\left[\frac{{M_{\rm pl}^2}}{2}F(\phi) R_H +\frac{1}{2}\eta^{\alpha\beta}(\partial_{\alpha}\phi)(\partial_{\beta}\phi) + V(\phi) \right] - {M_{\rm pl}}^2 R^{\mu\nu} F(\phi) 
\nonumber\\
& + & 
\frac{M_{\rm pl}{}^2}{2}  \eta^{\nu \rho} \left(\eta^{\lambda\mu}\eta^{\sigma\beta} - \eta^{\sigma\mu}\eta^{\lambda\beta} \right) \partial_{\rho}\left\{H_{\lambda\beta}\partial_\sigma F(\phi)\right\} - \frac{{M_{\rm pl}}^2}{2}\left[\eta^{\lambda \mu} \left(\partial^{\beta} \partial^{\nu}F \right) H_{\lambda \beta}  - \eta^{\lambda \beta}\left(\partial^{\mu} \partial^{\nu}F \right) H_{\lambda \beta}\right]
\nonumber\\
& + & 
\frac{{M_{\rm pl}}^2}{2} \eta^{\nu\beta}\left(\eta^{\lambda\mu}\eta^{\rho\sigma} - \eta^{\rho\mu}\eta^{\lambda\sigma} \right) \left[ \partial_{\rho}\left\{H_{\lambda\beta}\partial_{\sigma} F(\phi)\right\} - 3 \partial_{\sigma}\left\{H_{\lambda \beta}\partial_{\rho} F(\phi)\right\}\right]
\label{T111}
\end{eqnarray}
Here we have used the relation
\begin{eqnarray}
\left(\eta^{\lambda\mu}\eta^{\sigma\beta} - \eta^{\sigma \mu}\eta^{\lambda \beta} \right) \left\{\partial^{\nu}\partial_{\sigma} H_{\lambda\beta}\right\} F\left( \phi \right)
 & = & 2 {R_H}^{\mu\nu} F\left( \phi \right) + \left(\eta^{\mu\beta}\eta^{\nu \delta}\Box H_{\beta\delta} - \eta^{\nu\delta}
\partial^\lambda \partial_\mu H_{\lambda\beta} \right) F\left (\phi \right) \nonumber \\
& = & 2 {R_H}^{\mu\nu} F\left( \phi \right) + \partial_\sigma \left[\eta^{\nu\beta}\left( \eta^{\mu\lambda}\eta^{\rho
\sigma } -\eta^{\mu\rho}\eta^{\lambda
\sigma}\right)
\partial_\rho
H_{\lambda\beta}F\right] \nonumber \\
&& \qquad \qquad \qquad- \eta^{\nu\beta}\left( \eta^{\mu\lambda}\eta^{\rho
\sigma }- \eta^{\mu\rho}\eta^{\lambda
\sigma}\right)\partial_\rho(
H_{\lambda\beta}
\partial_\sigma F)
\label{manipulation}
\end{eqnarray}
obtained using  the substitution of $g_{\mu\nu}$ $H_{\mu\nu} $ in equation (\ref{1}) and (\ref{eqng}). 

The final step of our algorithm is to substitute $H_{\mu\nu}$ by $g_{\mu\nu}$ in (\ref{T111})
 to get
\begin{eqnarray}
\Theta^{\mu\nu} & = &-(\partial^{\mu}\phi)(\partial^{\nu}\phi) + \eta^{\mu\nu}\left[\frac{{M_{\rm pl}^2}}{2}F(\phi) R +\frac{1}{2}\eta^{\alpha\beta}(\partial_{\alpha}\phi)(\partial_{\beta}\phi) + V(\phi) \right] - {M_{\rm pl}}^2 R^{\mu\nu} F(\phi) 
\nonumber\\
& + & 
\frac{M_{\rm pl}{}^2}{2}  \eta^{\nu \rho} \left(\eta^{\lambda\mu}\eta^{\sigma\beta} - \eta^{\sigma\mu}\eta^{\lambda\beta} \right) \partial_{\rho}\left\{g_{\lambda\beta}\partial_\sigma F(\phi)\right\} - \frac{{M_{\rm pl}}^2}{2}\left[\eta^{\lambda \mu} \left(\partial^{\beta} \partial^{\nu}F \right) g_{\lambda \beta}  - \eta^{\lambda \beta}\left(\partial^{\mu} \partial^{\nu}F \right) g_{\lambda \beta}\right]
\nonumber\\
& + & 
\frac{{M_{\rm pl}}^2}{2} \eta^{\nu\beta}\left(\eta^{\lambda\mu}\eta^{\rho\sigma} - \eta^{\rho\mu}\eta^{\lambda\sigma} \right) \left[ \partial_{\rho}\left\{g_{\lambda\beta}\partial_{\sigma} F(\phi)\right\} - 3 \partial_{\sigma}\left\{g_{\lambda \beta}\partial_{\rho} F(\phi)\right\}\right]
\label{T1111}
\end{eqnarray}
and
identify the EMT (\ref{T1111}) to that of the original theory (\ref{QA}) in the adapted coordinates at the point P and then convert it to the general coordinates. Once this identification is done, $g_{\mu \nu}$ becomes the metric. 
The different terms on the right hand side of (\ref{T1111}) are in appropriate tensorial form suitable to be lifted to general coordinates. For instance,
\begin{equation}\frac{{M_{\rm pl}}^2}{2} \eta^{\nu \beta} \left( \eta^{\lambda\mu} \eta^{\rho\sigma} - \eta^{\rho\mu}\eta^{\lambda\sigma}\right) \partial_{\rho} \left(g_{\lambda\beta}\partial_{\sigma} F(\phi)\right) \to 
\frac{{M_{\rm pl}}^2}{2} g^{\nu \beta}\left(g^{\lambda\mu} g^{\rho\sigma} - g^{\rho\mu} g^{\lambda \sigma} \right)\nabla_{\rho} \left( g_{\lambda\beta} \nabla_{\sigma} F(\phi)\right).\label{cor}
\end{equation}
The right hand side of (\ref{cor}) is in covariant form and passes to the left hand side when the defining properties (\ref{g}), (\ref{delg}) and (\ref{del2g}) are used. Similar grneralization of the other terms is obvious, considering that $F$
is a scalar. One can now readily obtain the expression for the EMT in general coordinates, 
\begin{eqnarray}
\Theta^{\mu\nu}&=&-({\nabla^\mu}\phi)({\nabla^\nu}\phi)
+ {g^{\mu\nu}}[\frac{1}{2}{g^{\alpha\beta}}({\nabla_{\alpha}}\phi)({\nabla_{\beta}}\phi)+V(\phi)]
- {{M_{pl}^2}}F(\phi)
{G^{\nu\mu}} - {M_{\rm pl}}^2 g^{\mu\nu}
\Delta F  + {M_{\rm pl}}^2 \nabla^\mu \nabla^\nu F
\label{EMTF}
\end{eqnarray}
where $\Delta F = \nabla_{\mu} \nabla^{\mu} F\left( \phi \right)$ and $G^{\mu\nu}$ is the Einstein tensor. Note that
 metric compatibility is assumed. 

As can be checked directly, our algorithm, based on Noether theorem and dynamics of fields, leades to an EMT which is symmetric and covariantly conserved. Based as it is on the local conservation of energy-momentum, this EMT serves as the source of gravity {\it a la} Einstein.
Hence following the original spirit of general relativity, we write the equation of motion of non-minimally coupled quintessence model as 
\begin{eqnarray}
G^{\mu \nu} = -  \frac{1}{{M_{\rm pl}}^2}\Theta^{\mu\nu}
\label{EEQ}
\end{eqnarray} 
The correspondence with the Hilbert action principle must be investigated. First observe that if we put $F(\phi) = 0$ in (\ref{QA}) we get minimally coupled scalar field action. The corresponding EMT obtained by our algorithm is (see equation- (\ref{EMTF}))  
\begin{eqnarray}
\Theta^{\mu\nu}&=&-({\nabla^\mu}\phi)({\nabla^\nu}\phi)
+ {g^{\mu\nu}}[\frac{1}{2}{g^{\alpha\beta}}({\nabla_{\alpha}}\phi)({\nabla_{\beta}}\phi)+V(\phi)]
\label{EMT_no_F}
\end{eqnarray}
which is nothing but the EMT obtained by using (\ref{def}). Thus the connection of the usual EMT, obtained from (\ref{def}), as source of the gravitational field with the energy-momentum balance is properly elucidated by our algorithm. This is
a new result as far as we know.

The case of $F(\phi) \ne 0$, i.e., non-minimally coupled theories, is naturally more involved.
Such theories appear in different contexts, e.g., quantum corrections \cite{bd}, renormalization of classical theory \cite{cc}, in the string theoretic context \cite{sgc} and in the Scalar-tensor theories \cite{ref1}. Now a days such theories are intensly investigated in context of dark energy also. However, in the literature there are different approaches in identifying the EMT for a non-minimal theory.  To illustrate this, let us first write the equation of motion obtained by varying the metric of (\ref{QA}) 
\begin{eqnarray}
{M_{\rm pl}}^2 \left(1 - F\right) G_{\mu \nu} & = & \left(\nabla_{\mu}\phi \nabla_{\nu}\phi \right) - g_{\mu\nu} \left[\frac{1}{2}{g^{\alpha\beta}}({\nabla_{\alpha}}\phi)({\nabla_{\beta}}\phi) + V(\phi)\right]
+ {M_{\rm pl}}^2 g_{\mu\nu} \Delta F - {M_{\rm pl}}^2 \nabla_{\mu} \nabla_{\nu} F = \tilde{\Theta}{}_{\mu \nu}
\label{geqm}
\end{eqnarray}
One can keep the equation as it is \cite{H} and interprete $\tilde{\Theta}{}_{\mu \nu}$ as the EMT, which is clearly not covariantly conserved, thereby emphasising the deviation from the Einstein structure (\ref{EEQ}). The non-conservation has been criticised \cite{torre}. Alternatively, conserved EMT can be identified from (\ref{geqm}) by algebric manipulations which may take different courses \cite{ PB, torre} leading to EMTs with apparently different forms, e.g. 
\begin{eqnarray}
\Theta^{D}{}_{\mu \nu} & = &\frac{\tilde{\Theta}{}_{\mu \nu}}{\left(1 - F\right)} \label{T_D} \\
\Theta^{A}{}_{\mu \nu} & = & \tilde{\Theta}{}_{\mu \nu} + {M_{\rm pl}}^2 F G_{\mu \nu} \label{T_A}
\end{eqnarray}
Looking back at our equation (\ref{EMTF}) it is easy to see that it aggres with (\ref{T_A}). Thus our approach predicts a covariantly conserved EMT fro non-minimally coupled theories.


In this paper we have developed a novel algorithm to obtain the Energy momentum tensor (EMT) of scalar field theories coupled with gravity. The method rests on the corresponding theory where gravity is non dynamical. Adapted (Lorentzian) coordinates are chosen in the locally inertial frame where the first derivative of the metric vanishes
in an infinitesmal patch. We have then defined a field theory in the extended flat space that has the patch as overlap with the tangent space at P by the same action as the one obtained in the locally inertial frame. This theory has translation symmetry and the conserved Noether current gives an EMT. Identifying it as the EMT of the original theory in the locally inertial frame, we generalized the same to curved coordinates. 

 When the coupling of the scalar field with gravity is minimal, the standard form of the EMT is reproduced from the expression derived here. Non trivial results follow from the non minimally coupled scalar tensor theories, where incidentally different prescriptions of obtaining the EMT are available in the literature. It is gratifying to observe that our method leads to a covariant expression for the EMT which agrees with one of the widely used forms \cite{torre}. It is also apparent that the algorithm presented in this paper has a general applicability. 

\section*{Acknowledgement}
The authors acknowledge IUCAA for hospitality during their visit when a substantial part of the work was done. AS acknowledges the finantial support of DST SERB under Grant No. SR/FTP/PS-208/2012.

\end{document}